# Analysis and Evaluation of Real-time and Safety Characteristics of IEEE 802.11p protocol in VANET

*Hossein Ahmadvand*
*ahmadvand@ce.sharif.ir*

*Amir Hossein Jahangir*
*jahangir@sharif.ir*

*Ataollah Fatahi Baarzi*
*fatahibarzi@ce.sharif.ir*

*Department of Computer Engineering*
*Sharif University of Technology*
*Tehran, Iran*

*Abstract*—The need for safety in transportation systems has increased the popularity and applicability of Vehicular Ad-Hoc Networks (VANETs) in recent years. On-time reception and processing of alarms caused by possible accidents as well as the preventive actions have important roles in reducing human and financial losses in road accidents. In such cases, the performance of safety applications should be evaluated and guaranteed to show whether or not they can ensure the safety of humans and cars. In this paper, we analyze the behavior of Vehicular Ad-Hoc Networks by checking the real-time properties of the IEEE 802.11p protocol using a Colored Petri Net model. To analyze the performance of related standards, simulations are conducted using CPNTools. Standards from European Telecommunications Standards Institute (ETSI), and Vehicle Safety Communications (VSC) are evaluated in this research. We will show that such standards may not completely fulfill the safety requirements in particular situations.

**I.** INTRODUCTION

Wireless Access for Vehicular Environment (WAVE) is an approved amendment to the IEEE 802.11 standard and is currently used in vehicular communications. WAVE is required to be supported by Dedicated Short-Range Communications (DSRC) in Intelligent Transportation System (ITS) applications. There are three types of communications in VANETs: vehicle to vehicle (V2V), vehicle to infrastructure (V2I), and infrastructure to vehicle (I2V). All such communications are carried out in a 75-MHz-wide section of the 5.9 GHz band (5.85-5.925 GHz) [1, 2, 3].

Real-time constraint or equivantly deadline guarantee, are a very important issue in vehicular networks. In order to have a safe network, one should know how reliably and timely a (critical) transmission occurs. Receiving on-time warnings is crucial, especially for safety applications. First of all, a model of system or the communication protocol must be available then its real-time properties can be checked analytically. This model should be specified at the Application layer. If a warning is not received before its deadline specified by safety standards, the safety of the system (e.g. a human or a car) cannot be guaranteed. Hence a formal model is required for the safety properties to be validated by simulation.

In order to derive collision deadlines, we start from the Application Layer and use the Vehicle Safety Communications-Application (VSC-A) standard as the reference, and extract the communication latencies of the MAC layer from [4] and [5].

IEEE 802.11p describes a physical layer and Medium Access Control (MAC) that support channels for sending and receiving data. Other specifications for such a network is provided by the IEEE 1609.x stack protocol. IEEE 1609.4 defines multi-channel transmissions for 802.11p on the upper layer. This protocol divides the band into two channel types, each having an interval of 50 milliseconds: the Control Channel (CCH), and the Service Channel (SCH). CCH is used for transmitting short status messages and safety traffic massages [6]. When a safety message is to be sent during the time-slot of SCH, the message should be delayed until the time-slot of CCH starts. In IEEE 802.11p, the message will be further delayed because of the delays induced by the MAC and physical layers when framing the packets to be sent on the channel. To analyze the real-timeliness and the safety of VANET, we must describe all the deadlines and explain how they are defined and met in each application. In this paper, we analyze the seventh application of VSC (Pedestrian Crossing Information at Designated Intersections) as an example and report the results. What follows is the description of the application which is directly extracted from the final report of the Vehicle Safety Communications project-Task 3 [5]:

A. Application Definition

The application provides an alert for vehicles when there is the risk of a collision with a pedestrian or a child on a designated crossing.

B. Application Description

The presence of a pedestrian is detected through the infrastructure sensing equipment which includes the "walk" button that pedestrians press before crossing an intersection. A broadcast message containing the information regarding the pedestrian is transmitted from the roadside units to vehicles approaching the crossing area.

C. Communication Requirements

− One-way, Point-to-multipoint communication from the infrastructure to vehicles
− Transmission mode: periodic
− Minimum frequency (update rate): ~10 Hz
− Allowed latency: ~100 milliseconds
− Data to be transmitted and/or received: presence of a pedestrian
− Maximum required range of communication: ~200 m

Section II presents the related works. In section III, we describe the example scenario that we have simulated. The Petri Net model for the scenario is displayed in section IV. The Colored Petri Net is presented in section V. Section VI put forth our experimental results. The last section concludes the paper.

II. RELATED WORK

Our research has taken a different approach from the earlier ones. The researches around safety have paid no attention to modeling different applications at the Application layer. In the VSC-A project, the U.S. Department of Road Safety and the automotive companies such as General Motors, Ford, Honda, Toyota, and Mercedes-Benz have presented seven applications for seven accident scenarios, and in a pilot project, they have evaluated the reliability of V2V communications for the scenarios [5]. In [10], the performance of the IEEE 802.11p protocol has been evaluated. In this paper, the Physical and MAC layers are also taken into consideration and the analysis is carried out using the Petri Net modeling techniques so that we will have a formal time analysis.

So far, there have been limited researches on modeling the broadcast of messages in vehicular networks. For example, in [14], it is assumed that every vehicle has to send its beacons in the CCH period; otherwise, the period will expire and there is no queue for such beacons in the MAC layer. It is also assumed that every vehicle can only send one type of beacon in each CCH period. A mathematical event-based model is then presented and simulated by MATLAB and the MAC layer's performance evaluated. In [16], safety applications of VANET are analyzed precisely including speed and congestion. Markov model is used for evaluating the performance of the proposed model. Finally, performance of the proposed model is analyzed. According to the investigation carried out in [18], cellular network is considered as a main infrastructure in VANET. In addition, mixed infrastructure is utilized in this paper. In research [19] in order to enhance reliability of the network, another layer is added to the protocol. WAVE protocol is considered and Markov model are used to analyze the results. The main concept of paper [20] is about V2V communication. The combination of VANET and cloud computing is considered in the paper. The broadcasting delay is measured and reported accordingly. In [21] false warnings are declined significantly by implementing machine learning technology. As a result, the performance of VANET and probability of true reaction are increased. Colored Petri Net is utilized in [22] for modeling various states of VANET, and they are categorized based on criticality of safety. Resource allocation and task scheduling are discussed in paper [17] whereas Colored Perti Net is used as a modeling tool in the paper.

III. The EXAMPLE SCENARIO

The example scenario consists of a pedestrian that enters a road, and a vehicle that is moving on the road. We divide the time after the pedestrian enters the road into three intervals: the "Warning" interval, the "Perception" interval, and the "Reaction" interval.

A. The Warning Interval

During the Warning interval, an alarm is generated by the (sensor) network to inform the driver of the vehicle that a pedestrian is crossing the road. Assuming that the vehicle's velocity is constant, the distance traversed by the vehicle in this interval can be calculated as follows:

$$x_{Warning} = (v_0 \times t_{Latency}) \qquad (1)$$

Where $t_{latency}$ is the delay between the moment when the first alarm is sent by the network and the instant that the vehicle receives it, and $v_0$ is the vehicle' constant speed.

## B. The Perception Interval

In the *Perception* interval, the driver should react promptly and suitably to the alarm. This depends on the driver's age and mind state. The distance traversed by the car in this interval is:

$$x_{Perception} = v_0 * t_{Perception} \qquad (2)$$

where $v_0$ is the constant vehicle's velocity, and $t_{Perception}$ is the perception-reaction time, or the time elapsed between the instant the alarm is received and the moment when the driver starts to react. According to [12, 13], we assume that $t_{Perception}$ typically ranges from 0.7 seconds to 1.5 seconds.

## C. The Reaction Interval

The main factors affecting the *Reaction* interval are: the roads' wetness, and the vehicles' braking system. These two factors impacts the acceleration (or deceleration) of the vehicle during brakes. The distance traveled by the car during this interval is:

$$x_{Brake} = \frac{(v_0)^2}{2a} \qquad (3)$$

where $v_0$ is the constant vehicle's velocity, and $a$ is the acceleration of the vehicle before it immobilizes. We fix $a$ equal to $9 \, m/s^2$ when the road is dry, and $4 \, m/s^2$ when it is wet [12, 13].

## D. Total distance

The total distance traveled by the car is given by equation (4) as follows:

$$x_{total} = x_{Warning} + x_{Perception} + x_{Brake} \qquad (4)$$

If the driver is unable to stop the vehicle before traversing its total distance to the pedestrian, a collision will occur:

$$if \left( d - v_0 * (t_{Perception} + t_{Latency}) - \frac{(v_0)^2}{2a} \right) < 0 \rightarrow Collision \, Occures$$

In the above formula, $d$ is the distance between the vehicle and the pedestrian when the pedestrian enters the road, $v_0$ is the constant vehicle's velocity, $t_{Perception}$ is the driver's perception-reaction time, $t_{Latency}$ is the delay between the moment the first alarm is sent by the network and the instant when the vehicle receives it, and $a$ is the acceleration at during braking time.

## IV. PETRI NET MODEL

While other works such as [16] use models like Markov model for VANET, we believe that in order to model a system or a network protocol including VANET, Petri Net is a good candidate. There are several reasons which justify our choice:

*A.* Formal semantics:

A system specified by Petri Net has a clear and precise description, because it is formal and can be model checked mathematically or verified by simulation. Moreover, several enhancements have been made to classical Petri Net, such as colored, time, hierarchical, stochastic Petri Net, which makes it a universal formalism and tool for virtually all kinds of discrete event problems.

*B.* Graphical nature:

Petri Net is a graphical language being intuitive and easy to learn. The graphical nature eases communication with end-users.

*C.* Vendor independence:

Petri Nets provide a tool-independent framework for modeling and analyzing processes. Petri nets are not based on a software package of a specific vendor and do not cease to exist if a new version is released or when one vendor takes over another [14].

Petri Net is a graphical and mathematical modeling tool applicable to many systems. It is a promising tool for describing and studying information and processing systems characterized as concurrent, asynchronous, distributed, parallel, nondeterministic, and/or stochastic. As a graphical tool, Petri Net can be used as a visual-communication aid similar to flow charts, block diagrams, and networks. In addition, tokens are used to simulate the dynamic and concurrent activities in Petri Net. As a mathematical tool, we can set up state equations, algebraic equations, and other mathematical models governing the behavior of systems in Petri Net. Petri Net can be used both by practitioners and theoreticians between whom it provides a powerful medium of communication: practitioners can learn from theoreticians how to make their models more mathematical, and theoreticians can learn from practitioners how to make their models more realistic. [9]

The properties of Petri Nets lead to a vast diversity of its applications from network protocol and workflow modeling like [14], to business process modeling like [17]. Petri Net has four types of basic elements: Places, Transitions, Arcs, and Tokens. In this paper, it has been assumed that the events occur in places. For modeling the described scenario we should use places and transitions to map events and states respectively. Figure 1 illustrates the Petri Net model of the seventh application of the VSC standard. Tables 1 and 2 describe the transitions and places of the model respectively.

As described in Table 1 and Table 2, P0 is the initial state of system. In P1 the pedestrian enters the street. After T1 is fired, the system goes to such a state that the pedestrian is crossing, the traffic light is green for him/her and red for cars. If the presence of pedestrian is determined by the system, in P4 the Warning signals (network packets) will be created, otherwise the systems goes to unsafe state. While the system is not in unsafe state and warning signals are created, in P6 that network packets which carry the warning signals are broadcast through the network and sent to drivers. Therefore in P7, the packets are received by drivers. In P9, if the driver reacts correctly, the system goes to safe state and no collision occurs, else a collision would occur between the car and pedestrian.

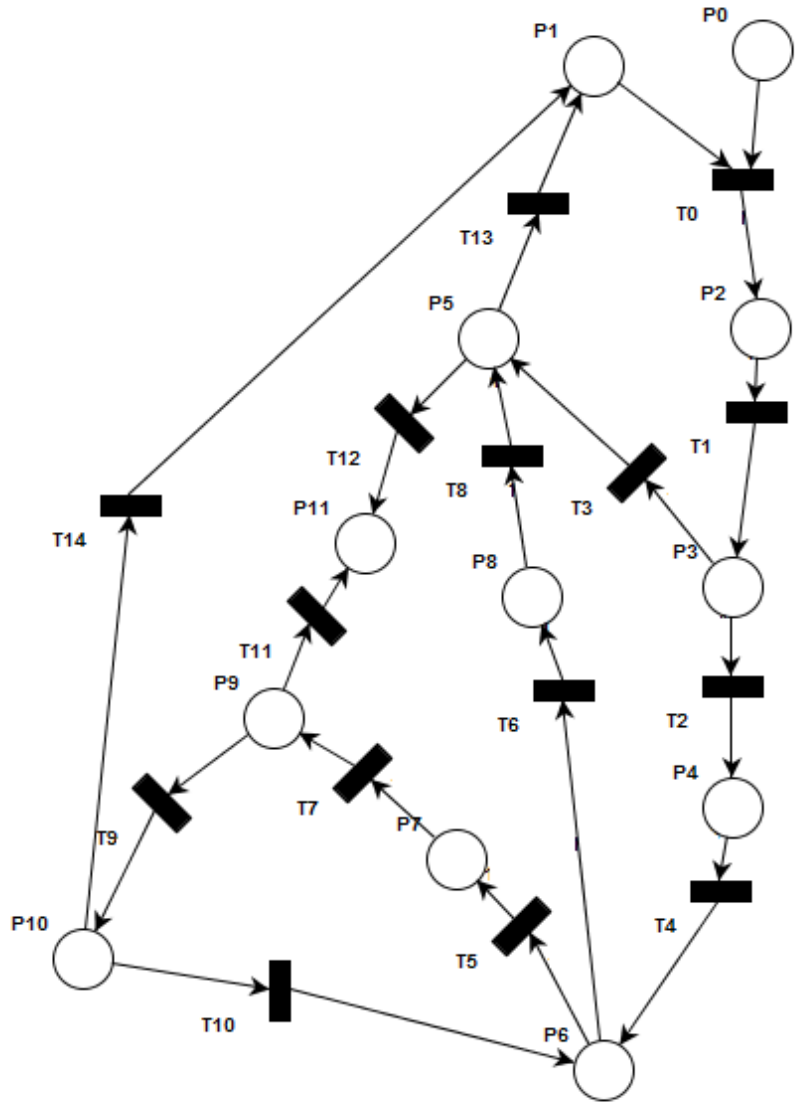

**Figure 1- The Petri Net model of the example scenario**

## Table 1- Description of the transitions in the model

| Transition | Description |
|---|---|
| T0 | This transition is used to exit the initial state of the scenario |
| T1 | The pedestrian, for whom the traffic light is green, is crossed |
| T2 | The Pedestrian presses the Crossing button |
| T3 | The Pedestrian has not pressed the Crossing button |
| T4 | Broadcast by the car or by the Road Side Unit (RSU) |
| T5 | The packet is received correctly with a probability of $(1 - P_r)$ |
| T6 | The packet is not received correctly with a probability of $P_r$ |
| T7 | The total delay of driver's reaction |
| T8 | The transition will be fired if packet loss ratio is higher than a threshold |
| T9 | The transition will be fired if the driver reacts correctly |
| T10 | Pedestrian crossing is not complete and the scenario will continue |
| T11 | The transition will be fired if the driver's response is incorrect and a collision occurs |
| T12 | Incorrect reaction is made by the driver when the state is unsafe. This will lead to a collision. |
| T13 | Correct reaction is made by the driver when the state is unsafe. This will avoid a collision. |
| T14 | The model starts over again. |

## Table 2- Description of the places in the model

| Place | Description | Place | Description |
|---|---|---|---|
| P0 | Initial state | P6 | The alert is broadcast |
| P1 | Presence of a pedestrian | P7 | Warning message is received by the driver |
| P2 | Pedestrian enters the street | P8 | Packet loss occurs |
| P3 | While the pedestrian is crossing, the traffic light is green for him/her and red for cars | P9 | The driver reacts |
| P4 | Warning signals are sent to drivers | P10 | Safe state |
| P5 | Unsafe state | P11 | Collision |

Figure 2 shows the inputs and outputs of the model. The model provides a description of the system and its behavior with respect to its inputs. The presence of tokens in places show the active states of the system. For example, the presence of a token in the collision place shows that a collision has occurred.

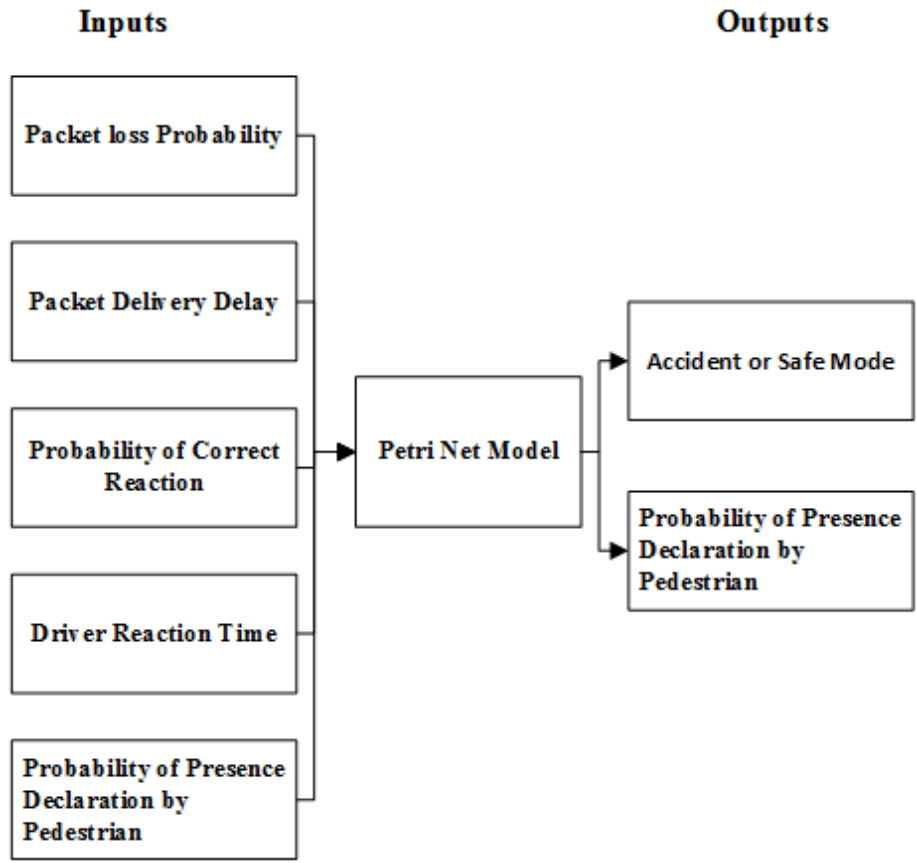

Figure 2- The inputs and outputs of the Petri Net model

V. The SAFETY COLORED PETRI NET MODEL

In this section, we describe how we have built a formal model of the safety protocol of vehicular environments using CPNTools.
In CPNTools, we use some variables to specify the time or deadlines of events for related places and transitions. Each transition is fired if all its prior places have the required number of tokens and the time assigned to them has elapsed. All alarms are mapped to tokens, and vehicle movement states are represented by places. If alarms are not received on-time, the state of the vehicle might change; for example, if the alarm showing the presence of a pedestrian is not received before its deadline, the state of the system might change to unsafe.

The places in this scenario have been defined as "TIxSTRING" in which "TI" is a "colset" (type) defined as a type of "int timed" allowing us to observe the timing behavior of tokens. "STRING" is a colset defined as a type of "string" that lets us show token information in the simulator interface. Variables such as "PPPD" (stands for Probability of Pedestrian Presence Detection), "PLP" (stands for Packet Loss Probability) and CRP ( stands for Correct Reaction Probability) are used when we want certain tokens to be removed from the model (not received). For example, if the type of a place to which 19 (out of 20) tokens are assigned is "PPPD" or "CRP", the probability that a token can pass the place is 95%. We divided our CPN model into five parts which are depicted in Figures 3 to 7.

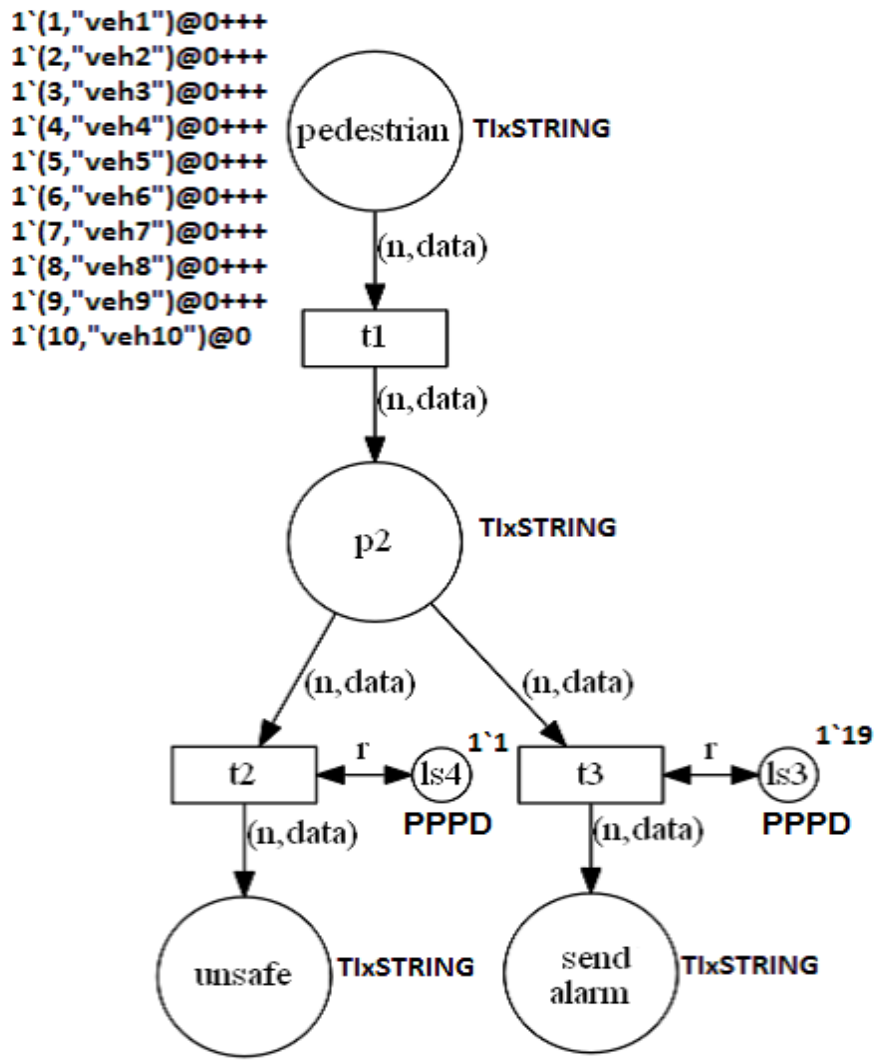

**Figure 3- The start state of system**

As illustrated in Figure 3, the system starts when the pedestrian enters the road. If "t2" is fired, it means that the pedestrian has not announced its presence by pressing the Pedestrian Crossing button, and if "t3" is fired, it means that the pedestrian has done so. The probability of "t2" being fired is 5% after which the system goes to the "unsafe" place; the probability of "t3" being fired is 95% after which an alarm should be sent with the system going into the "send alarm" place.

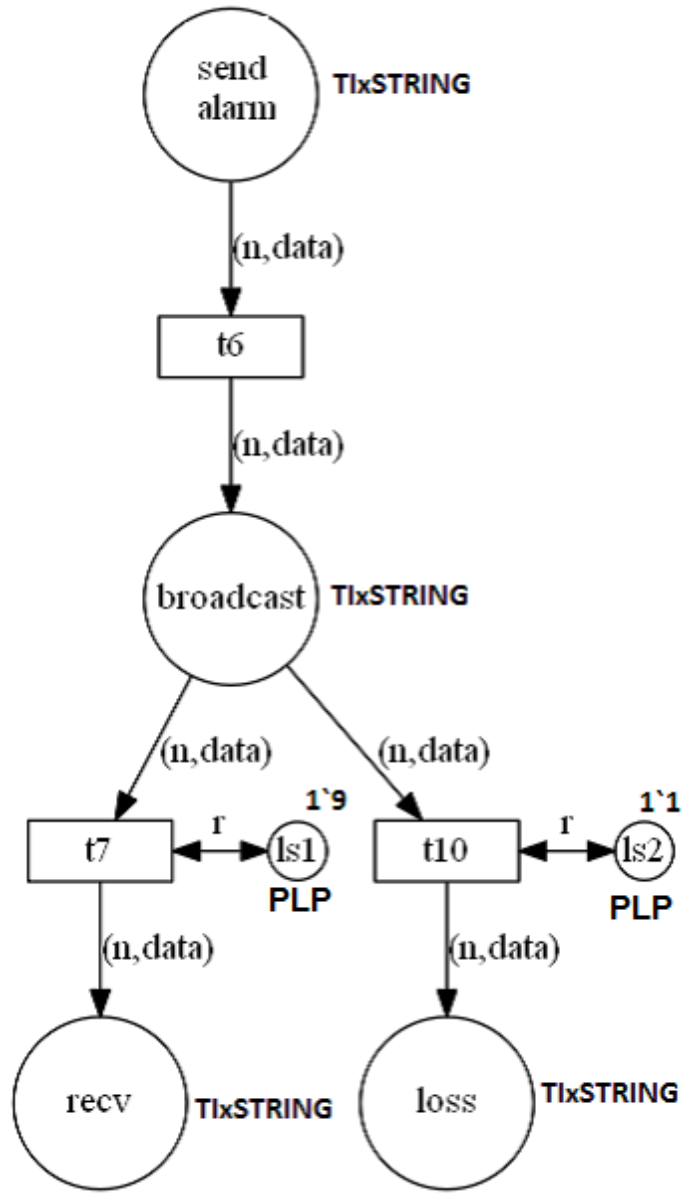

**Figure 4- Broadcasting the alarm (network packet) to the network**

As shown in Figure 4, in order to inform the presence of a pedestrian an alarm (network packet) should be broadcast to the network, when the system is in the "send alarm" place. In the "broadcast" place, the alarm is sent to other vehicles. As shown in the same figure, the packet loss probability is 10%, and the packet reception probability is 90%.

If the packet is lost, the system will go to the "unsafe" place and if the vehicle receives it, the system will go to the "reaction" place which means that the driver should reacts to the presence of the pedestrian in the road. (Figure 5).

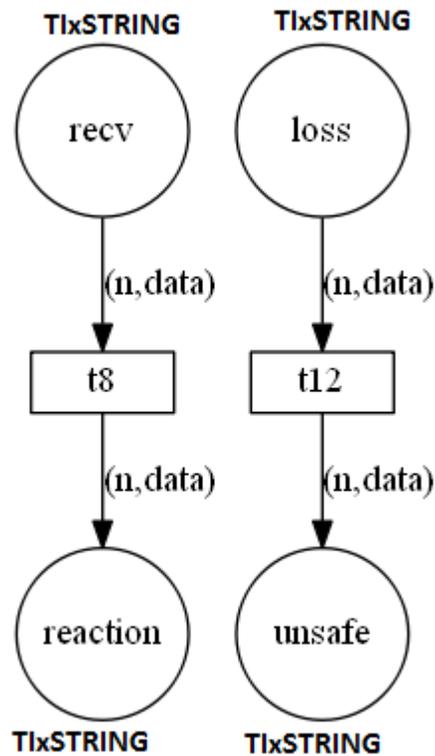

**Figure 5- Driver reaction to the presence of the pedestrian.**

The reaction flow is presented in Figure 6. The probability of an incorrect reaction is 5% and that of a correct one is 95%. In the "reaction" place, if the driver reacts incorrectly the system will go to the "collision" place, and if a correct reaction occurs, the system will be safe and will go to the "pedestrian" place which means that the system starts over again.

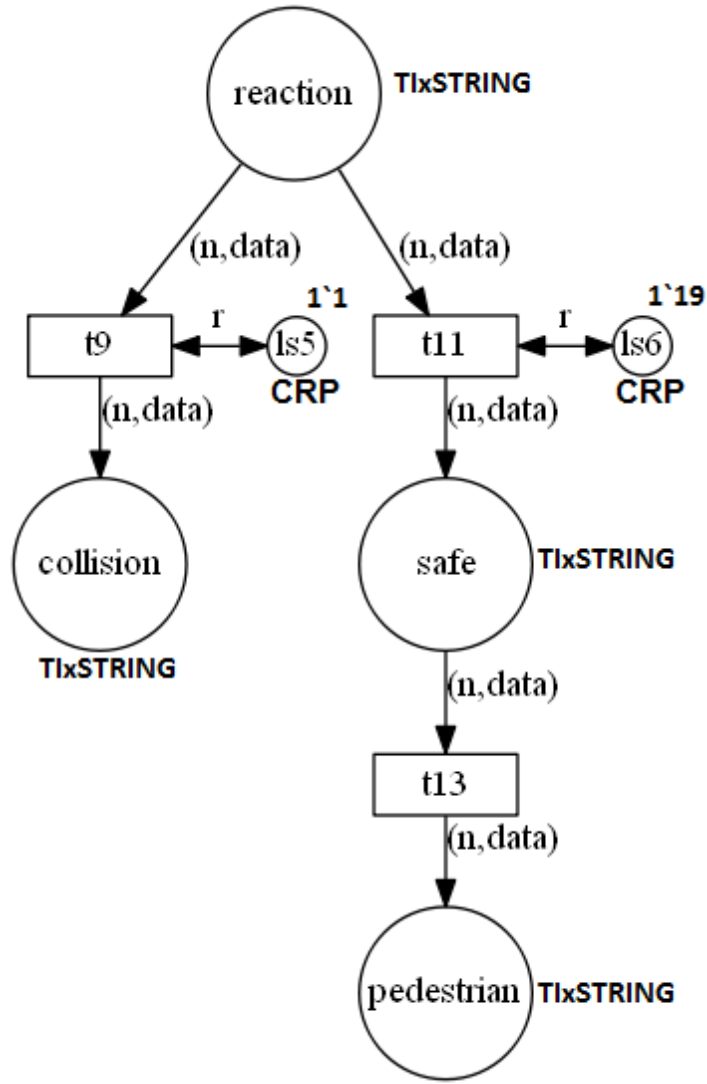

**Figure 6- Flow of correct or incorrect reaction**

Finally, in Figure 7, when the system is in the "unsafe" place, either a collision occurs or the pedestrian completely crosses the road with the system going to the "safe" place.

It is worth noting that by losing the first alarm message, the system state will not go to the collision place. In order for such a transition to occur, more than one message should be lost. The number of lost messages prompting the system's state to transit to the collision place can vary depending on the scenarios.

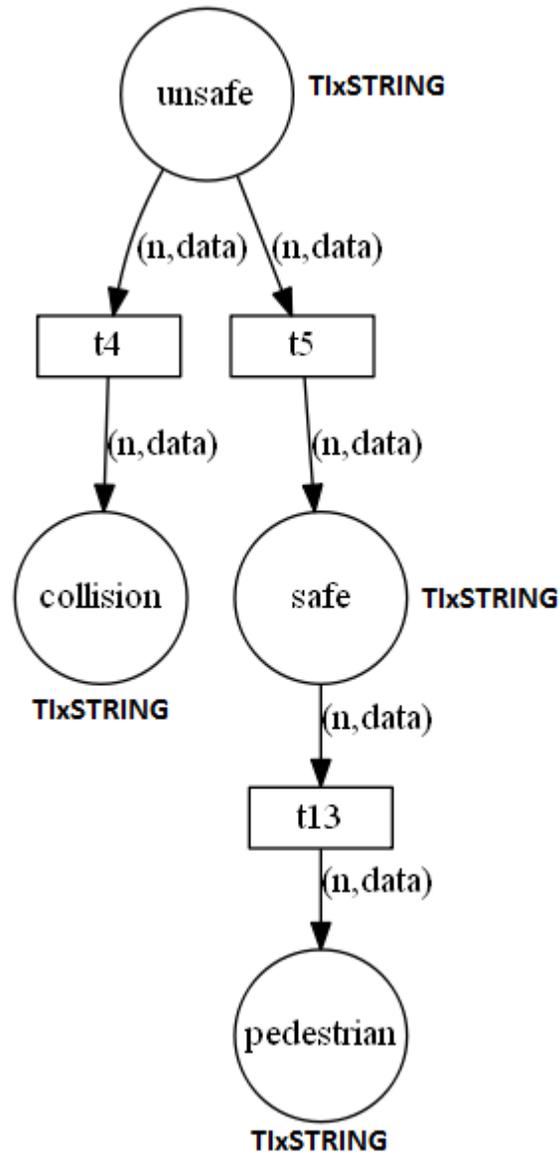

**Figure 7- Unsafe state of system**

In this model, we need to estimate some parameters such as packet loss probability, latency of network and broadcast, correct reaction probability, and pedestrian declaration probability. Our goal is to analyze the VSC standard over IEEE 802.11p in pessimistic scenarios. For example, by simulating the network, we measure the worst latency. We use such information to estimate the delay of the network and broadcast in the transition "t7". We also measure the packet loss probability for different numbers of vehicles by the same manner. We have assumed that packets are lost only at the beginning of the intervals; hence, the scenario can be considered as pessimistic since this yields to maximum delay in reception of the alarms by the driver. We have used previous studies to determine the delay and packet loss percentages [7, 8]. These papers examine the packet loss and the reception delay in different conditions. In addition, we have simulated the scenario in OMNeT++ and validated the results. The simulation parameters are given in Table 4.

**Table 4- Simulation parameters in OMNeT++**

| Parameter | Value |
| --- | --- |
| Number of vehicles | 5 − 138 |
| Packet rate(packet/second) | 10 |
| Packet size(bit) | 2048 |
| Simulation arear($m^2$) | 450 ∗ 450 |
| Channel type | Wireless |
| Mac Layer | 802.11p |
| Traffic type | CBR |

Figure 8 shows the worst case latency in the network which has been extracted from our simulation results. We used such results to evaluate the worst case delay. As shown in Figure 8, the delay of packet arrivals increases when the number of vehicles increases.

Our results show that the delay becomes maximum when there are 57 vehicles because the network becomes saturated at this point. From this point onwards, the delay decreases because packets are lost with higher probabilities resulting in a less congested network.

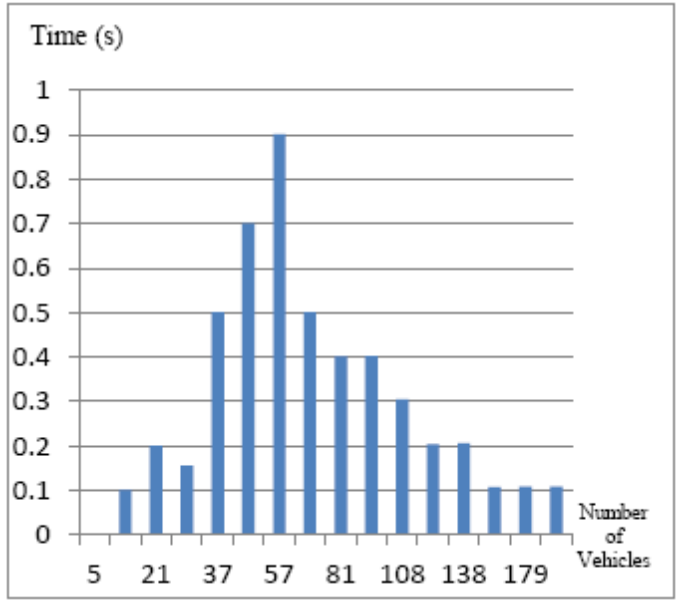

**Figure 8- The worst case latency of the network for the pedestrian scenario**

Figure 9 shows the packet loss ratio in the pedestrian scenario. In this figure, an increase in the number of vehicles augments the packet loss ratio as well.

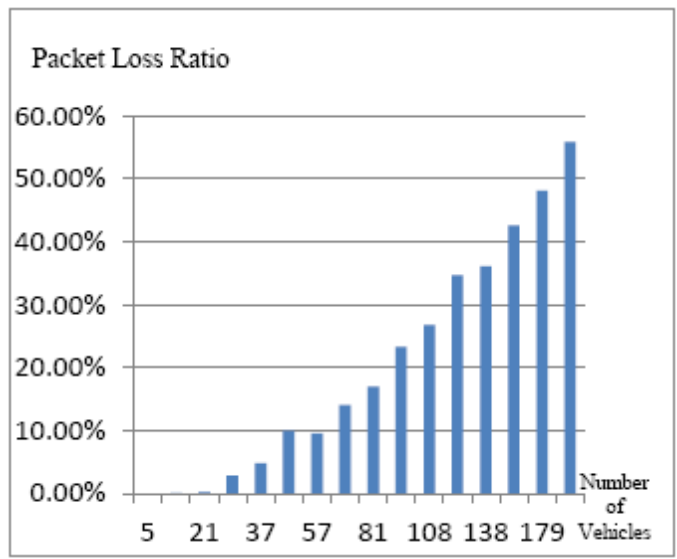

**Figure 9- Packet loss ratio in the pedestrian scenario**

We have tried to adopt a pessimistic situation: we assume that the first alarm packets are lost (with an arbitrary probability) in each interval, and the packets that are successfully transmitted will be received with maximum delay. The observations show that when we rely only on the network for detecting the presence of pedestrians while the driver is somehow unable to detect them, the scenario may fail from a safety standard point of view due to packet loss and delay in reception (for instance, in the case of the presence of some fog or similar anomalies). In this study, we analyze only the safety applications that send packets in one hop without any retransmission.

VI. EXPERIMENTAL RESULTS

In the VSC standard, the warning is sent to the driver when the distance between the driver and the pedestrian becomes less than 200 meters. In this case, if the driver does not stop the vehicle in a timely manner, a collision will occur. The most important parameters affecting the specified distance are network latency, perception reaction time, and braking acceleration. We have assumed the constant values of 1 second and $4\,m/s^2$ for $tpr$ and $a$ respectively. As observe in Figure 10, the driver cannot stop his vehicle in less than 200 meters in certain conditions and speeds resulting in a collision. The figure shows that in some cases, even when the number of vehicles is very low (5 to 27), the driver cannot control the vehicle until a distance of 200 meters is travelled. The problem becomes more serious when the number of vehicles increases, as shown in the figure. Hence, it can be noticed that, regardless of the number of vehicles, the VSC protocol suffers from safety problems.

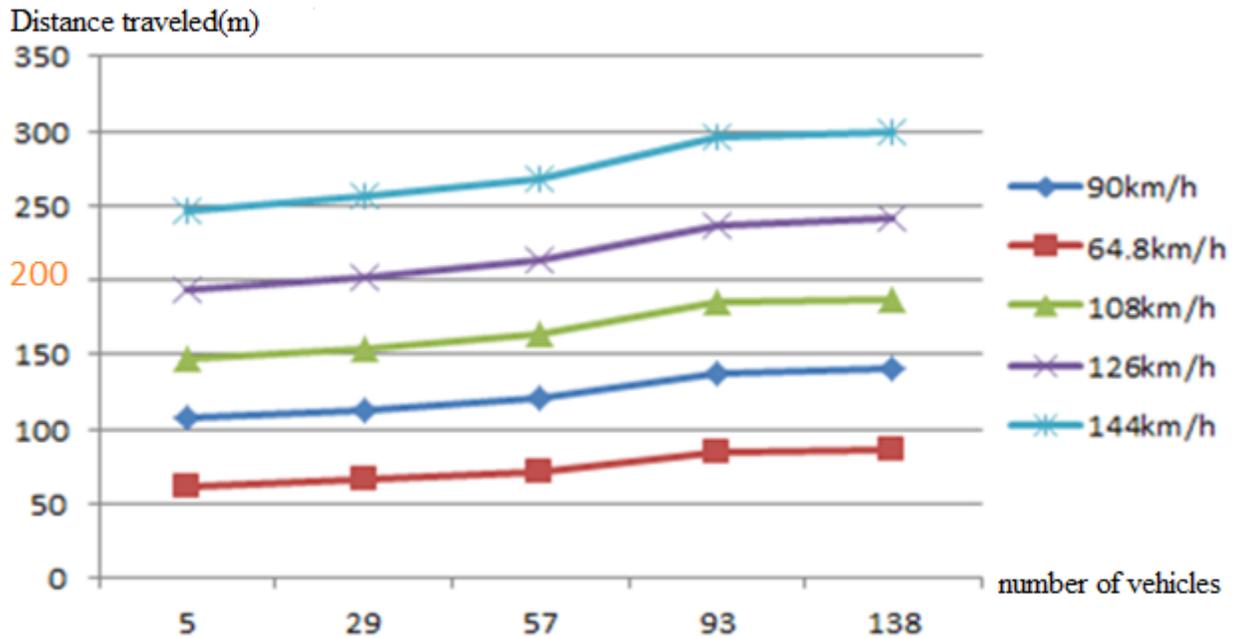

**Figure 10- The distance travelled for different number of vehicles and speeds**

VII. CONCLUSIONS

In this paper, some challenging and vulnerable parts of VSC-A were modeled by Colored Petri Net. The modeled scenarios show that in some cases, the standard cannot guarantee the safety of the driver, the car or the pedestrians, and some enhancements or corrections should be made at MAC layer or application layer of the protocol. The protocol analyzed in this study is used for vehicular network; therefore, it must guarantee the safety of humans and vehicles using it whereas such deficiencies could not be tolerated. By referring to our analytical and simulation results, it can be noticed that the standard is not totally safe for some scenarios such as pedestrian cross, and corrections should be applied to it. For instance, the warning distance must be incremented or the maximum speed of driving should decrease.

As future works, other challenging and critical parts of the safety standards related to VANET can be modeled and verified by using both formal and practical methods in order to detect new possible vulnerabilities. Combining safety and commercial applications to evaluate their impact on safety are a good step forward in this field.